\documentclass[epj]{svjour}
\usepackage{graphicx}
\begin{document}

\title{Critical Behavior of a Three-State Potts Model on a Voronoi Lattice}

\author{F. W. S. Lima\inst{1,2} \and  U. M. S. Costa\inst{1} 
\and M. P. 
Almeida\inst{1}
\and J. S. Andrade Jr.\inst{1,3}}

\institute{Departamento de F\'{\i}sica, Universidade Federal do Cear\'a,
60455-760 Fortaleza, Cear\'a, Brazil \and
Universidade Estadual Vale do Acara\'u, Sobral, Cear\'a, Brazil\and
PMMH-ESPCI, 10 rue Vauquelin, 75231 Paris CEDEX 05, France}
\abstract{
We use the single-histogram technique to study the critical behavior of 
the
three-state Potts model on a (random) Voronoi-Delaunay lattice with size 
ranging from 250 to 8000 sites. We consider
the effect of an exponential decay of the interactions with the distance,
$J(r)=J_0\exp(-ar)$, with $a>0$, and observe that this system seems to have
 critical exponents $\gamma$ and $\nu$ which are 
different from the respective exponents 
of the three-state Potts model on a regular square lattice. However, the ratio
$\gamma/\nu$ remains essentially the same. We find numerical evidences 
(although not conclusive, due to the small range of system size) that 
the specific heat on this random system behaves as a power-law for $a=0$ and 
as a logarithmic divergence for $a=0.5$ and $a=1.0$
}

\maketitle 
\section{Introduction}
The randomness in the lattice of statistical spin models has been 
studied in order to access the effect of impurities and dilutions over their 
critical behavior. It was conjectured by Harris \cite{Har1974} that the sign 
of the critical exponent of the specific heat, $\alpha$, determines whether
the system is affected or not by such a randomness. For positive values of 
$\alpha$, the impure system should have a critical behavior different from 
the one of the pure system. For negative values of $\alpha$, on the other 
hand, the critical behavior of the system should be the same for both cases.
In the marginal case of $\alpha=0$, one should not be able to draw any 
conclusion about changes in the system. 

The pure ferromagnetic three-state Potts model has $\alpha=1/3$, hence,
from the Harris criterium we expect to find a different behavior with a 
random interaction system. However, Picco \cite{Picco} used this model with 
two types of interactions ($J_0=1$ and $J_1=1/10$) randomly and equally
distributed ($p=0.5$) and did not find significant differences from the pure
case. In another study \cite{Jae-Kown}, a quenched random interaction between
nearest neighbors has been introduced, where the coupling factor $J_{ij}$ 
for each pair $i,j$ is selected from two positive values $J$ and $J'$ with 
respective probabilities equal to $p$ and $1-p$. It was found that the 
exponent $\eta(=2-\gamma/\nu)$ does not depend on the disorder length, 
while $\nu$ and $\gamma$ vary continuously with this type of disorder, 
satisfying the concept of weak universality proposed by Suzuki \cite{Suzuki}.
The finite size scaling behavior of the Potts model with three and four states
on regular square lattice has been studied by Kim and Landau \cite{Kim1998}, 
where they found that for $q=4$ the multiplicative logarithmic correction is 
insufficient to correct the dominant terms.

In the present work, we investigate the effect of quenched disorder on
the three-state Potts model using a two dimensional Voronoi-Delaunay network. 
This kind of disordered lattice exhibits a random coordination number that 
varies from 3 to $\infty$, depending on the number of sites. In addition, 
the distance $r$ between nearest neighbors changes randomly from site to 
site. This geometrical feature is incorporated in our model by using 
a coupling parameter that is an exponential function of $r$,
$J(r) \propto \exp(-ar)$, with $a>0$. In a previous work, we observed that the
two-dimensional ferromagnetic Ising model with the same coupling mechanism
and the same lattice displays critical behaviors for $\alpha$, $\beta$ and $\nu$
that are independent of the value of $a$ and are the same as those of the pure
system \cite{Lima}. Here we extend this analysis to the three-state
ferromagnetic Potts model. 

\section{Model and Simulation}
The Voronoi construction or tessellation for a given set of points is defined
as follows.
$N$ points are randomly placed in a square with side lengths equal to 
$N^{1/2}$. Then, for each point we determine the polygonal cell consisting
of the region closer to this point than to any other. The points whose
cells share an edge are considered neighbors. The dual lattice (Delaunay's) 
is composed by  the vertices of these polygons.
The lattice obtained by linking the neighbor sites is the Voronoi network.

The Hamiltonian for the three-state ferromagnetic Potts model is given by
$$
-KH= \sum_{<i,j>}J_{ij}\delta(\sigma_i-\sigma_j),
$$
where $K=1/k_B T$, $T$ is the temperature, $k_B$ is the Boltzmann constant and
$\delta$ is the Dirac's delta. The summation is performed over all
pair of neighbors in the Voronoi's network and the spins can take values in the 
set $\{1,2,3\}$. We assume that 
$$
J_{ij}=J(r_i-r_j)=J_0\exp(-a|r_i-r_j|),
$$ 
where $J_0$ is a dimensionless and positive interaction factor, $r_i$ and 
$r_j$ are the position vectors of the sites $i$ and $j$, respectively, 
and $a\ge 0$ 
is a model parameter. 

For each grid size, we use the single-cluster algorithm 
\cite{Wolf} to simulate the system in 
the vicinity of the phase transition, i.e., with $K=K_{C_{\mbox{max}}}$, 
to determine the system finite size behavior. 

The value of $K_{C_{\mbox{max}}}$ is determined in the following way.
We  find the first estimate $K_0$ as the point of maximum specific heat 
obtained in the curve  (specific heat against $K$) constructed by simulations
with $K$ varying. 
The value of $K_0$ is used to
perform from $R=10$ to $20$ independent simulations with $10^6$ MC steps each, 
from which, using the single
histogram algorithm \cite{Ferrenberg}, 
we construct a series of curves of $C$ against $K$. 
We then average over all of these curves in order to find the 
specific heat curve as
\begin{equation}
C(K)=(1/R)\sum_{i=1}^{R}C^{(i)}(K),
\end{equation}
where $C^{(i)}(K)=K^2 N(<e^2>_K -<e>_K^2)$ is computed using the 
single-histogram with the results of simulation $i$. Finally, we obtain,
from the average curve,
the maximum value of the specific heat, $C_{\mbox{max}}$, and its abscissa
$K_{C_{\mbox{max}}}$.

In order to estimate the ratios $\beta/\nu$ and $\gamma/\nu$, we use 
the fact that the magnetization at the inflection point and the 
susceptibility should scale, respectively, as
\begin{equation}\label{eq2}
|<M>|_{\mbox{inf}}= L^{-\beta/\nu}f(tL^{1/\nu})\propto L^{-\beta/\nu},
\end{equation}
\begin{equation}\label{eq3}
\chi_{\mbox{max}}(L)= \chi(K_{C_{\mbox{max}}}(L),L) = AL^{\gamma/\nu},
\end{equation}
and that 
\begin{equation}\label{eq4}
\left| {d \over dK}<|M|>\right|_{\mbox{max}} = L^{-\beta/\nu+ 1/\nu} 
f'(tL^{1/\nu}) \propto L^{(1-\beta)/\nu}.
\end{equation}
Hence, the logarithmic derivatives should scale as
\begin{equation}\label{eq7}
\left|{d \over dK}\mbox{ln} <|M|>\right|_{\mbox{max}}= 
\left| {1\over |M|}{d \over dK}<|M|>\right|_{\mbox{max}} \propto L^{1/\nu},
\end{equation}
and 
\begin{equation}\label{eq8}
\left|{d \over dK}\mbox{ln} <|M^2|>\right|_{\mbox{max}}= 
\left| {1\over |M^2|}{d \over dK}<|M^2|>\right|_{\mbox{max}} \propto L^{1/\nu},
\end{equation}
and so we can get a good estimate of $\nu$ from their scaling behavior.
Finally, from finite size scaling arguments we can also predict that 
the specific heat should scale as
\begin{equation}\label{eq9}
C_{\mbox{max}}(L)= B_0+ B_1\mbox{ln}L.
\end{equation}

\section{Results and Conclusions}
We study the critical behavior of the Potts model for three values of $a$
($a=0.0, 0.5$ and $1.0$). For each value of $a$, we apply the finite size 
scaling
technique \cite{Fisher} together with the single-histogram algorithm.
We perform the same procedure for systems with different number of sites 
$N=250$, 500, 1000, 2000, 4000, and 8000. 
The critical temperature for infinite size system is estimated 
by using the fourth-order magnetization (Binder) cumulant and
we find the critical values $K_C=0.607$, $1.035$ and $1.959$ and $U^*=0.606$, 
$0.615$ and $0.623$, corresponding to $a=0.0$, $0.5$ and $1.0$, respectively.

In Fig.~1 we show a log-log plot of $M$ against $L$($=N^{1/2}$) for $a=0.0$,
$0.5$ and $1.0$. By linear fitting each of these plots and using Eq.~(\ref{eq2}), 
we obtain $\beta/\nu=-0.133$, $-0.118$ and $-0.106$,
respectively. The errors in these measurements are in order of ten percent, 
which is true also for the plots of the other figures. This is a consequence
of the small range of the system size.

In Fig.~2 we show the plot of $\chi_M$ against $L$ also in logarithmic scale,
from which we obtain the values of the exponent $\gamma/\nu=1.764$,
$1.751$ and $ 1.754$ for $a=0.0$, $0.5$ and $1.0$, 
respectively. As in \cite{Jae-Kown}, we observe that both $\gamma$ and $\nu$ 
vary with $a$ while the ratio $\gamma/\nu$ remains essentially the same.

Figures 3 and 4 show the plots of the logarithmic derivatives (\ref{eq7}) 
and (\ref{eq8}) against $\mbox{ln}L$. The slopes of the curves
produce the estimates for $1/\nu$, from which we get the values
$\nu_1=0.840$, $0.934$ and $1.060$ and 
$\nu_2=0.841$, $0.934$ and $ 1.061$ for
$a=0.0$, $0.5$ and $1.0$ respectively.

In Fig.~5 we show the plots of $C_{\mbox{max}}$ versus $\mbox{ln}L$ for
values of $a=0.0$, $0.5$ and $1.0$. We observe that the curves for $a=0.5$ 
and $a=1.0$ can be well fitted by a straight line while the curve for $a=0.0$
can not. The least-squares fits to data give the estimates
$B_0=-0.818$ and $0.956$ and $B_1=1.464$ 
and $ 0.506$ for $a=0.5$ and $1.0$ respectively.  
Figure 5 also contains the exponential fittings of $C_{\mbox{max}}$ 
versus $L$. From this figure, we can conjecture that, in the case of $a=0.0$, 
the specific heat behaves like a power-law of $L$. In order to provide 
a quantitative support for this argument, we present in Table~1 the sum 
of the square errors obtained by the linear regressions (first row) and 
by the exponential fittings (second row) of $C_{\mbox{max}}\times\mbox{ln}L$.

Table~2 contains our numerical results in a condensed form. From the 
analysis of this data, we conclude that the randomness, which is introduced 
here through the geometry of the Voronoi-Delaunay lattice, changes the critical
behavior of the system. This can be clearly observed by the change 
in the values of $\gamma$ and $\nu$. The value of $\beta$ is not affected
by the randomness. The analysis of the behavior of the exponent $\alpha$ 
is not conclusive since it is difficult, due to the limited number of 
data points, to distinguish the trend (power-law or logarithmic) 
in the curve of $C_{\mbox{max}}$ versus $L$. As already mentioned, 
the fittings shown in Fig.~5 provide some indication that the specific 
heat is a power-law of $L$ for $a=0.0$, and varies as a linear function 
of $\mbox{ln}L$ for $a=0.5$ and $1.0$. This conjecture needs to be 
checked, however, for values of $L>8000$.

\noindent{\bf Acknowledgements}

This research was partially supported by CNPq and Funcap (Brazilian agencies).

\newpage

\begin{table}[htb]
\caption{Total square residuals of linear and exponential 
regressions of $C_{\mbox{max}}\times\mbox{ln}L$ data}
\begin{center}
\begin{tabular}{|c|c|c|c|}\hline
$a$ & 0.0 & 0.5 & 1.0 \\ \hline \hline
Linear &  1.262949e-01 & 6.017198e-03 & 7.392830e-04 \\
Exponential & 6.369170e-03 & 1.474494e-02 & 3.294589e-03\\
\hline 
\end{tabular}
\end{center}
\end{table}

\begin{table}[htb]
\caption{Theoretical and computed values of the critical exponents}
\begin{center}
\begin{tabular}{|c|c|c|c|c|c|}\hline
& $\beta$ & $\gamma$ & $\nu$  & $\beta/\nu $ & $\gamma/\nu$ \\ \hline \hline
2d-Potts($q=3$)  & $1/9$& $13/9$& $5/6$ & $2/15$& $26/15$ \\ 
\hline
$a=0.0$ & 0.112 & 1.482 & 0.840 & 0.133 & 1.764 \\
$a=0.5$ & 0.110 & 1.635 & 0.934 & 0.118 & 1.751\\
$a=1.0$ & 0.112 & 1.859 & 1.060 & 0.106 & 1.754\\ 
\hline
\end{tabular}
\end{center}
\end{table}
 
\centerline{Figure Captions}

Fig.~1 -- Logarithmic plots of magnetization ($\mbox{ln}(M)$) versus 
$\mbox{ln}(L)$ for $a=0.0$ (circles), $0.5$ (squares) and $1.0$ (diamonds).

Fig.~2 -- Logarithmic plots of susceptibility ($\mbox{ln}(\chi)$) versus 
$\mbox{ln}(L)$ for $a=0.0$ (circles), $0.5$ (squares) and $1.0$ (diamonds).

Fig.~3 -- Plots of $\mbox{ln} \left(\left| {d \over dK}\mbox{ln}<|M>
\right|_{\mbox{max}}
\right)$ versus $\mbox{ln}(L)$ for $a=0.0$ (circles), $0.5$ (squares) 
and $1.0$ (diamonds).

Fig.~4 -- Plots of $\mbox{ln} \left(\left|{d 
\over dK}\mbox{ln} <|M^2|>\right|_{\mbox{max}}\right)$ versus 
$\mbox{ln}(L)$ for $a=0.0$ (circles), $0.5$ (squares) and $1.0$ (diamonds).

Fig.~5 -- Plots of $C_{\mbox{max}}$ versus $\mbox{ln}(L)$ for $a=0$ (circles),
$0.5$ (squares) and $1.0$ (diamonds). Also shown are the linear (solid lines) 
and exponential (dashed-lines) fittings to the data.

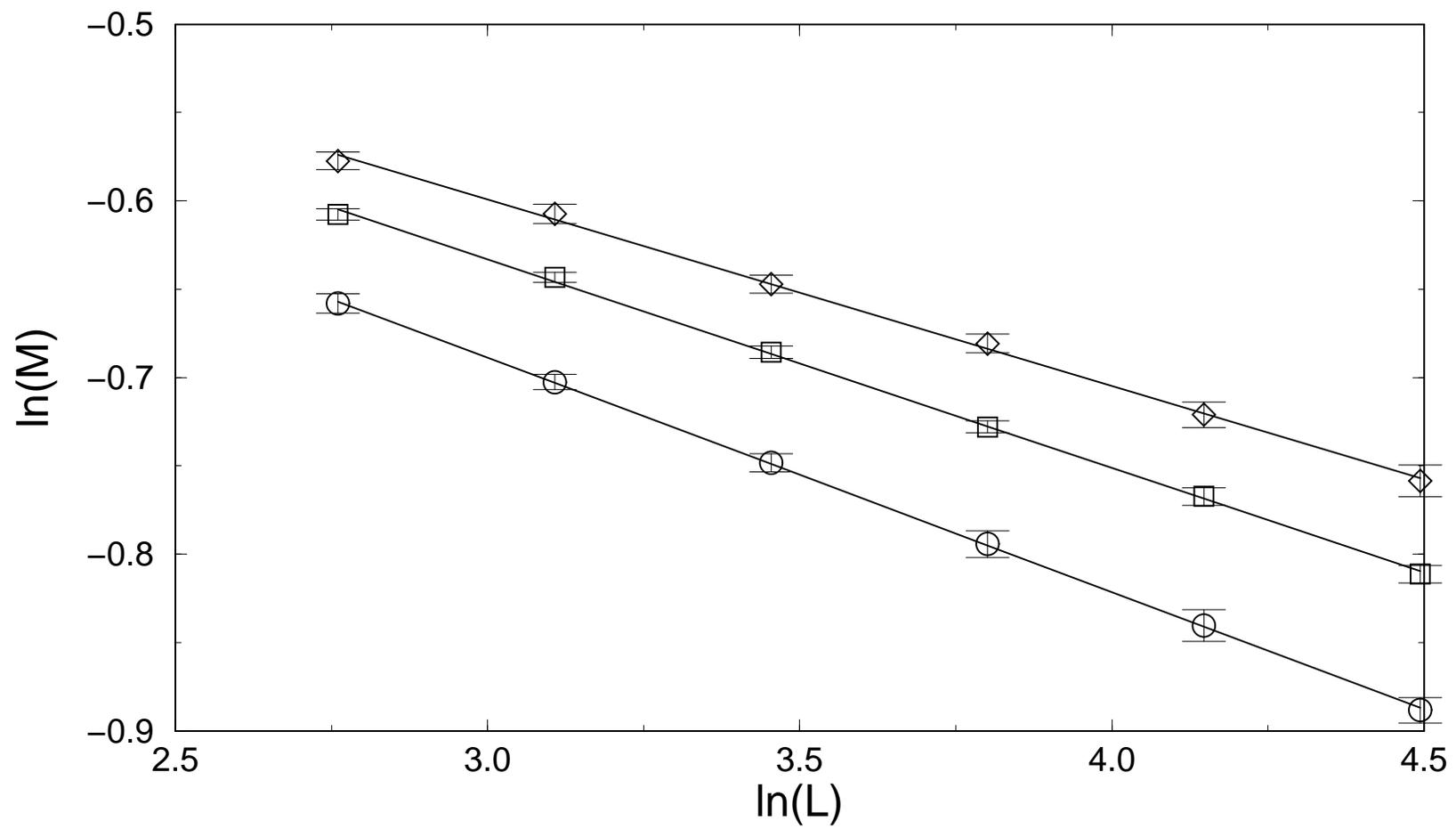
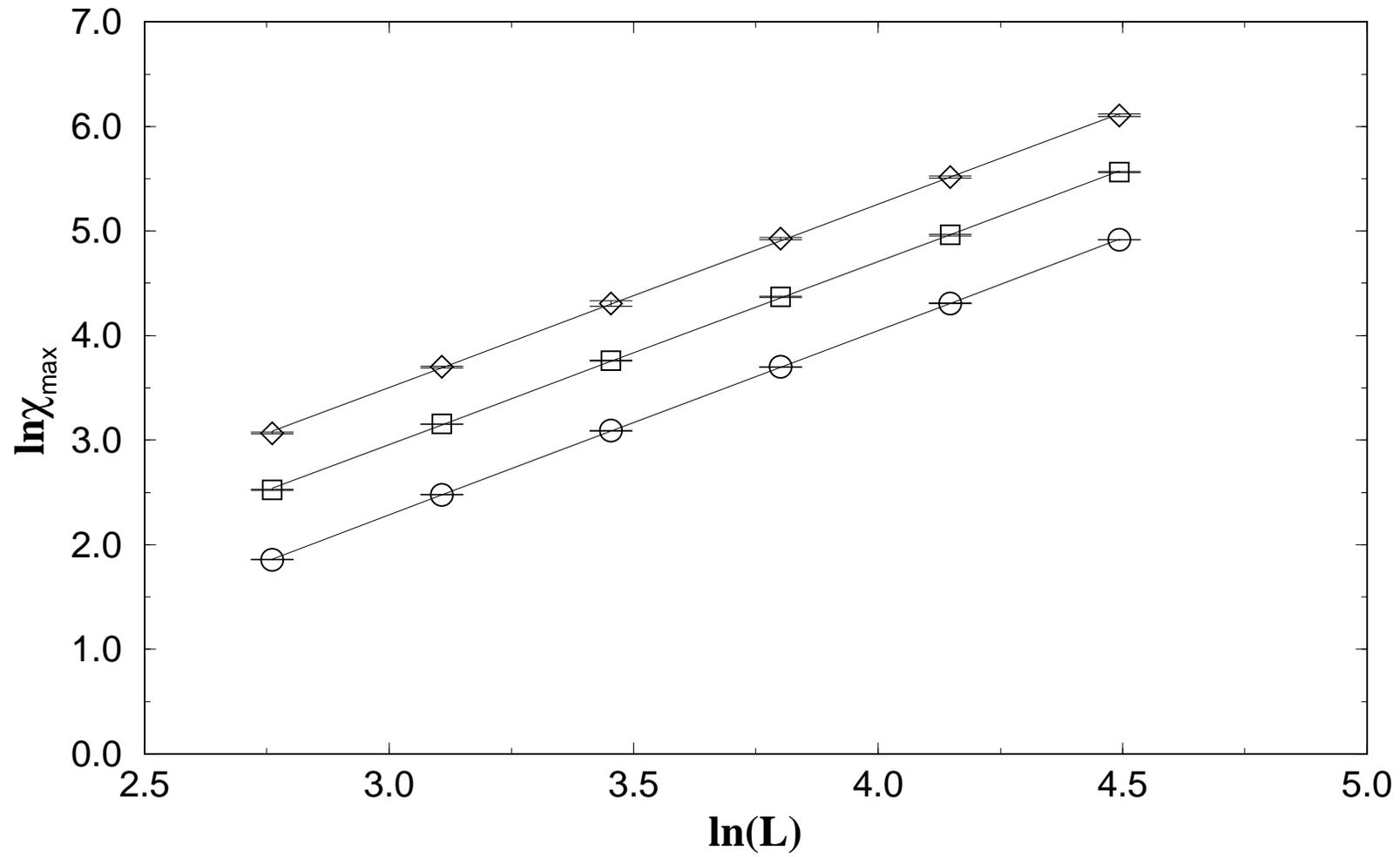
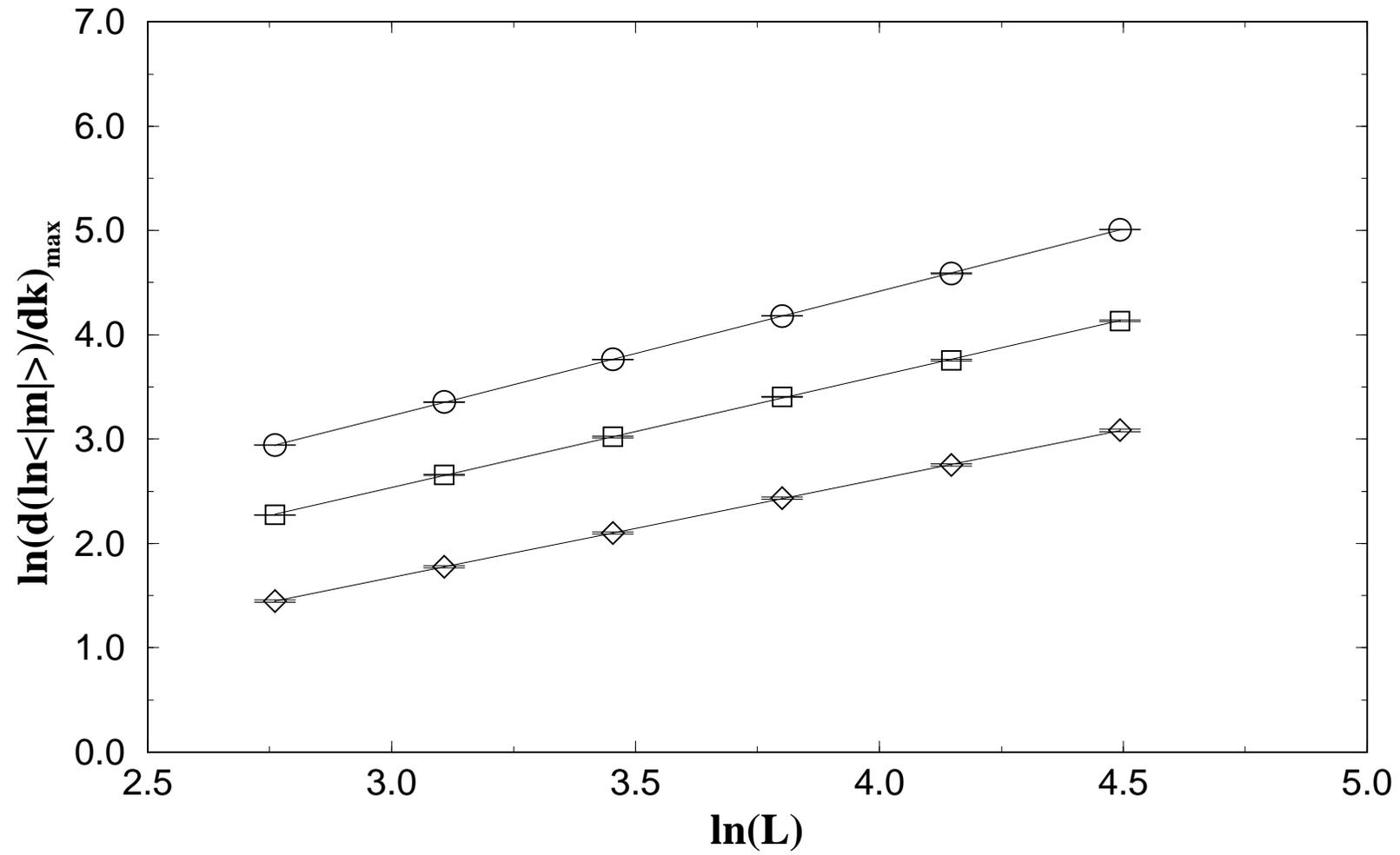
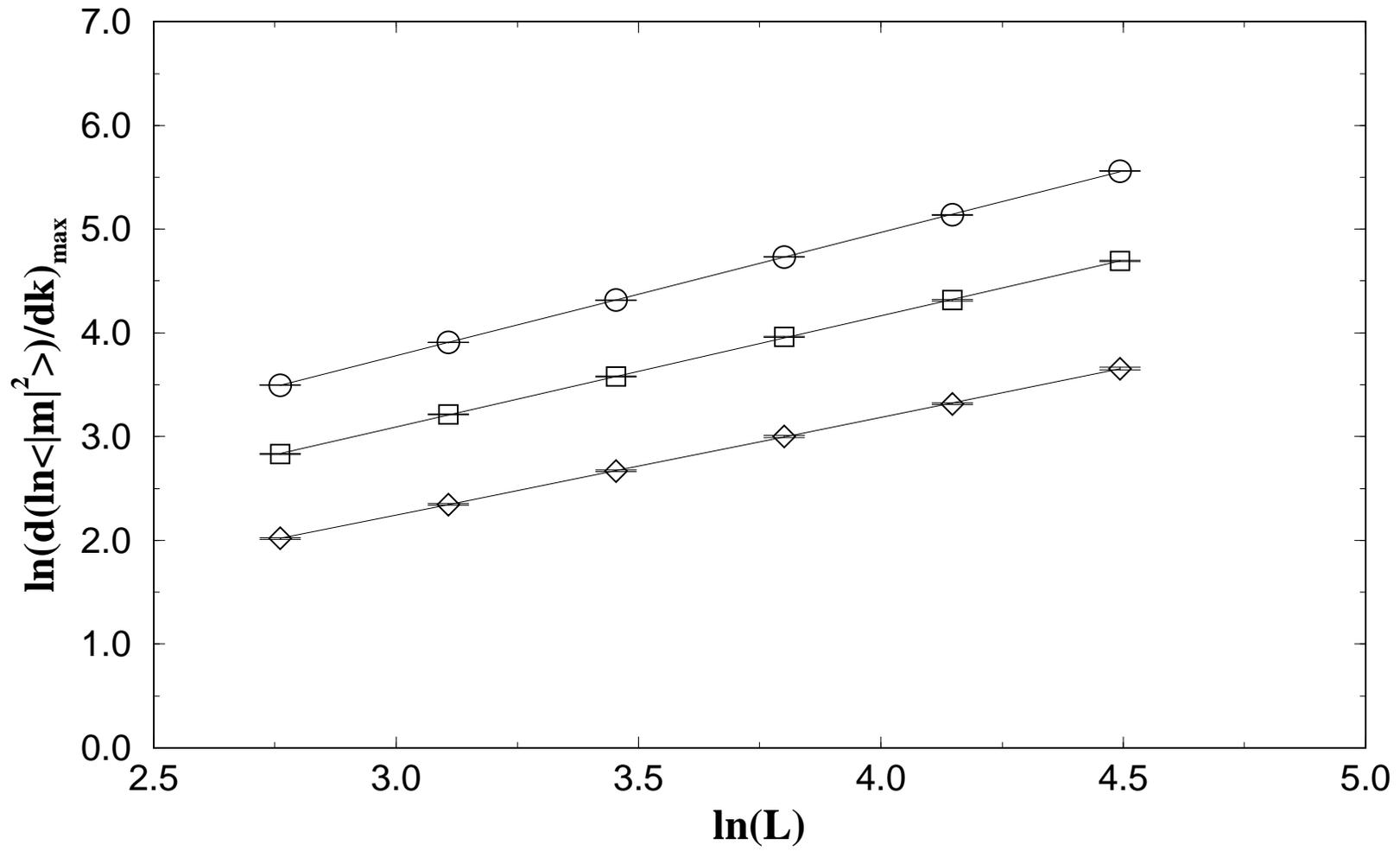
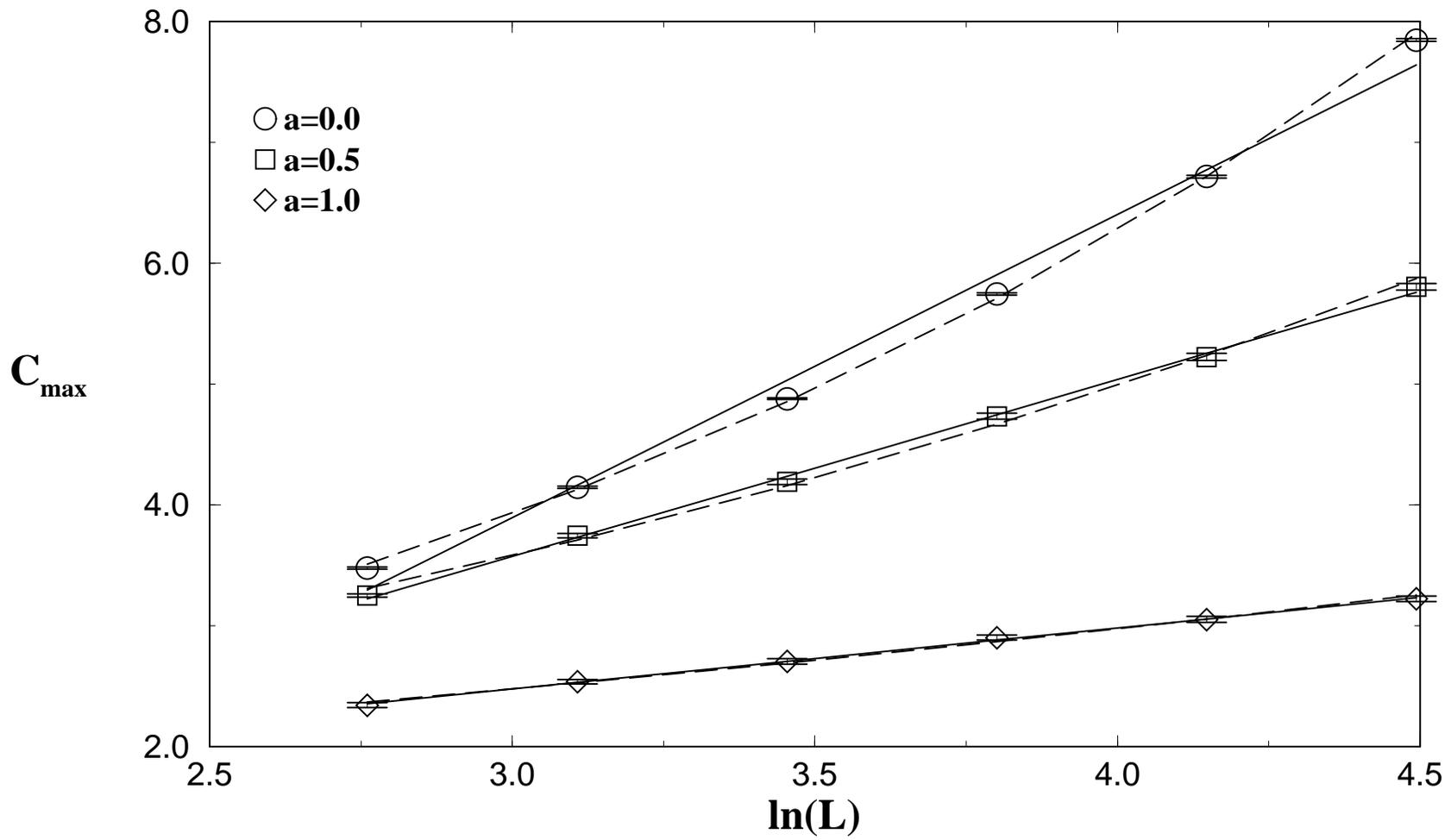
\end{document}